\documentclass[12pt]{article}
\usepackage{epsfig}

\newcommand{\blank}[1]{\ \\[#1 ex]}

\newcommand{\CC}{
{{\mathchoice {\setbox0=\hbox{$\displaystyle\rm C$}\hbox{\hbox
to0pt{\kern0.4\wd0\vrule height0.9\ht0\hss}\box0}}
{\setbox0=\hbox{$\textstyle\rm C$}\hbox{\hbox
to0pt{\kern0.4\wd0\vrule height0.9\ht0\hss}\box0}}
{\setbox0=\hbox{$\scriptstyle\rm C$}\hbox{\hbox
to0pt{\kern0.4\wd0\vrule height0.9\ht0\hss}\box0}}
{\setbox0=\hbox{$\scriptscriptstyle\rm C$}\hbox{\hbox
to0pt{\kern0.4\wd0\vrule height0.9\ht0\hss}\box0}}}} }

\newcommand{\supscript}[1]{$^{\mbox{\footnotesize#1}}$}

\newcounter{ctr1}
\newcounter{ctr2}
\newcounter{ctr3}
\newcounter{ctr4}

\newtheorem{rem}{{\em Remark}}           % Individual numbered remarks

\begin{document}

\centerline{ \Large Optimal design of PID controllers using the QFT method}
\centerline{      }
\centerline{\large
   A.\ C.\ Zolotas\supscript{1} and G.\ D.\ Halikias\supscript{2}}
      \footnotetext{\supscript{1}
       Department of Electronic and Electrical Engineering,
       Control Group, Loughborough University, Loughborough, LE11 3TU, UK}
      \footnotetext{\supscript{2}
      Department of Electronic and Electrical Engineering,
      University of Leeds,
      Leeds LS2 9JT, UK}
\blank{2} \noindent

{\bf Abstract: } An optimisation algorithm is proposed for designing
PID controllers,
which minimises the asymptotic open-loop gain of a system, subject
to appropriate robust-stability and performance QFT constraints.
The algorithm is simple and can be used to automate the loop-shaping
step of the QFT design procedure. The effectiveness of the method is
illustrated with an example.


\section{Introduction}

Many practical systems are characterised by high uncertainty which makes
it difficult to maintain good stability-margins and performance properties
for the closed-loop system. There are two general design methodologies
for dealing with the effects of uncertainty: (i) {\it Adaptive control}, in
which the parameters of the plant (or some other appropriate structure)
are identified on-line and the information obtained is then used to
``tune'' the controller, and (ii) {\it Robust control}, which typically
involves a ``worst-case'' design approach for a family of plants
(representing the uncertainty) using a single fixed controller.

{\it Quantitative Feedback Theory} (QFT) is a robust-control method developed
during the last two decades which deals with the effects of uncertainty
systematically. It has been sucessfully applied to the design of both
SISO and MIMO systems, while the theory has also been extended to the
nonlinear and the time-varying case. In comparison to other optimisation-based
robust control methods, QFT offers a number of advantages. These
include, (i) the ability to assess quantitatively the ``cost of feedback''
\cite{HOR}, \cite{HS1}, \cite{HS2}, (ii) the ability to take into account
phase information in the
design process (which is lost if, e.g. singular values are used as the
design parameters), and , (iii) the ability to provide ``design
transparency'', i.e. clear tradeoff criteria between controller
complexity and feasibility of the design objectives. Note that (iii)
implies in practice that QFT often results in simple controllers
which are easy to implement.

For the purposes of QFT, the feedback system is normally described by
the two-degrees-of-freedom structure shown in Figure 1. In this diagram,
$G(s)$ is the uncertain plant, $K(s)$ is the feedback controller and
$F(s)$ is the pre-filter. The objective is to design $K(s)$ and $F(s)$
so that the output signal $Y(s)$ tracks accurately the reference signal
$R(s)$ and rejects the disturbance $D(s)$, despite the presence of
uncertainty in $G(s)$.

\begin{figure}[htb]
   \epsfig{file=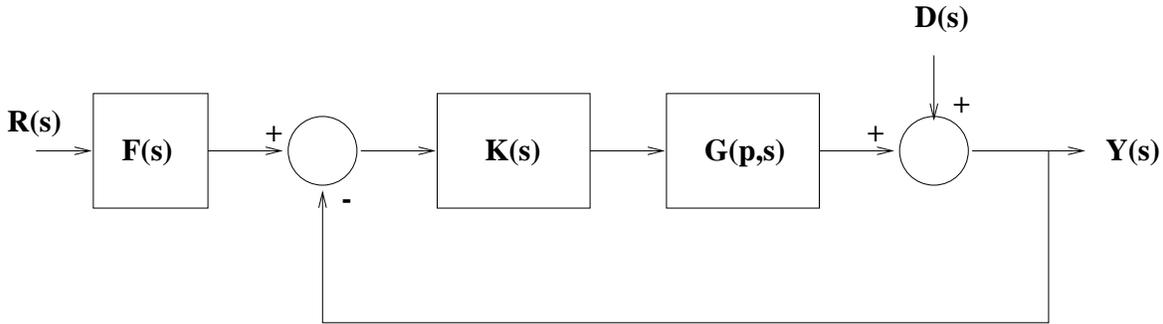,scale=0.8}
   \caption{Two degree of freedom feedback system}
   \label{fig:1}
\end{figure}

The uncertain plant $G(s)$ is assumed to belong to a set, $G(s) \in
\{ G(p,s) : p \in {\cal P} \}$, where $p$ is the vector of uncertain
parameters; these are assumed to be fixed but unknown, and to take
values in ${\cal P}$. The uncertain plant is first
translated in the frequency domain (using a discrete grid of frequencies
$\omega_1, \omega_2,...,\omega_N$, typically chosen to cover adequately
the system's bandwidth), resulting in $N$ ``uncertainty templates''
${\cal G}_i= \{ G(p,j \omega_i ) : p \in {\cal P} \}$, $i=1,2,...,N$.
The tracking specifications are given in the form of upper ($B_u (
\omega_i )$) and lower ($B_l( \omega_i )$) bounds in the frequency
domain, usually based on simple second-order models to represent
appropriate underdamped and overdamped conditions. A similar procedure
is followed to formulate the disturbance-rejection objectives of the design:
In this case, the magnitude of the sensitivity function $| S(p,j \omega_i)|=
| (1+G(p,j \omega_i) K(j\omega_i ) )^{-1} |$
is required not to exceed an appropriate bound $D_u(\omega_i )$ for
all $i=1,2,...,N$ and every $p \in {\cal P}$. Note that the disturbance
bounds are defined independently of the tracking bounds due to the presence
of the prefilter, $F(s)$.

Next, the tracking and disturbance-rejection specifications are translated
to certain conditions on the nominal open-loop frequency response $L_o(j
\omega)=
G_o(j\omega) K(j\omega)$ where $G_o(j\omega )$ denotes the nominal plant,
defined for {\it any} $p \in {\cal P}$. Consider first the tracking bounds.
For each frequency $\omega_i$, $i=1,2,...,N$, the tracking specifications
will be satisfied if and only if,
\begin{equation}
{\rm max}_{p \in {\cal P}} \Delta
\left| \frac{G(p,j\omega_i)K(j\omega_i)}
{1+G(p,j\omega_i)K(j\omega_i)} \right|_{\rm db}
\leq \delta(\omega_i) :=B_u(\omega_i) |_{\rm db}-B_l(\omega_i) |_{\rm db}
\end{equation}
i.e. if and only if the maximum variation in the closed-loop gain as
$p$ varies over the set ${\cal P}$,
at each frequency $\omega_i$ does not exceed the maximum allowable spread
in the specifications $\delta(\omega_i)$. This is
because, (i) the uncertainty associated with $K$ is assumed negligible, and
(ii) the actual closed-loop gain can always be adjusted to its required
level at each frequency via the scaling action of the prefilter. For each
frequency $\omega_i$, the open-loop gain (for each open-loop phase $\phi$)
at which
condition (1) is met with equality, defines the so-called ``Horowitz
template'' $h(\omega_i,\phi)$; this is the minimum open-loop gain
necessary to achieve the specified robust tracking specifications
at frequency $\omega_i$. In practice, each ``Horowitz template'' is calculated
via a bisection algorithm over a phase grid, and the equality condition in
(1) is satisfied within a given tolerance. In total, we have $N$ Horowitz
templates, one for
each frequency of interest $\omega_i$. A similar procedure will result in
$N$ robust disturbance-rejection contours, $d_i( \omega_i,\phi )$, $i=1,2,
...,N$, which define the minimum open-loop gain required to achieve robust
disturbance rejection (i.e. disturbance rejection for every $p \in {\cal P}$).
Clearly, to achieve robust tracking and robust disturbance-rejection
{\it simultaneously}, the open-loop frequency response must satisfy,
\begin{equation}
| L(j \omega_i) | \geq f_i (\omega_i, \phi) := {\rm max} \{ h_i(\omega_i, \phi)
, d_i(\omega_i, \phi) \}
\end{equation}
for all $i=1,2,...,N$, where the maximum in (2) is calculated pointwise in
$\phi={\arg}L(j\omega_i)$. The contours $f_i (\omega_i,\phi)$ will be
referred to as the {\it robust-performance} bounds.

In addition to robust-performance objectives, the closed-loop system
should also be robustly stable, i.e. the closed-loop transfer function
$F(s)G(p,s)K(s)(1+G(p,s)K(s))^{-1}$ should be stable for every $p \in
{\cal P}$. Since $F(s)$ will be designed stable, robust stability
can be inferred from the number of encirclements around the $-1$ point
by the open-loop frequency response $L(p,j\omega)=G(p,j\omega)K(j\omega)$,
$p \in {\cal P}$.
In practice, a more severe constraint is imposed on $L(p,j\omega )$: To
establish a minimum amount of damping for the (nominal) closed-loop
system, the nominal open-loop response is constrained not to enter an
$M$-circle of an appropriate value. Under the assumption of parametric
uncertainty\supscript{1} in the plant, the uncertainty templates of
$G_o(j\omega )$ at high frequencies approach a vertical
line in the Nichols chart. Hence, to ensure that the specified minimum
amount of damping is maintained
at high frequencies despite the presence of uncertainty, the $M$
circle is translated downwards by a specific amount, resulting in the
so-called ``universal
high frequency boundary'', which should not be penetrated by the
nominal open-loop response \cite{HS1}. This contour will be denoted by
${\cal B}$ in the sequel.

\footnotetext{\supscript{1}
The method can be modified to also take into account unstructured high frequency
uncertainty.}

\section{Optimal Design of PID controllers}

After constructing the contours $f_i (\omega_i,\phi)$ and ${\cal B}$ on the
Nichols' chart, QFT normally proceeds with the design of the
feedback controller $K(s)$. This normally involves frequency-shaping
of the nominal open-loop frequency response $L_o(j\omega)$, so that
it does not penetrate the the ${\cal B}$ contour and
$|L_o(j\omega_i)|$ lie on or above the robust-performace bounds
$f_i(j\omega_i,\phi)$ for each $\omega_i$.
At this stage, the designer normally follows a phase-lead/lag
compensation approach which involves a considerable trial-and-error element,
and can be cumbersome, especially if
the number of specified frequencies is large and the specifications
are tight. If the specifications can not be achieved, the design
objectives are assumed to be infeasible. In this case, the specifications
are normally relaxed and the design is repeated. If the specifications
are feasible, the best design is considered to be the one for which the
specifications are met as tightly as possible. This is in order to avoid the
possibility of ``overdesigning'' the system using unnecessarily large
gains/bandwidth, which can result in measurement noise amplification
and potential instability due to parasitics and high-frequency
unmodelled dynamics \cite{HOR}, \cite{HS1}. A compromise between controller
complexity and a ``tight'' design has to be made in many cases.

In this section we present a simple algorithm for designing PID
controllers which are optimal in the QFT sense. PID (or ``three-term'')
controllers are widely used in industry because they are simple and
easy to tune. Our algorithm can be used to provide an adequate QFT
design, or as the first step for designing a more complex controller \cite{BH}.
In addition, the algorithm can easily tackle a large number of constraints
and can be also be applied to multivariable systems, using the standard
QFT approach \cite{YH}, \cite{MAC}.

A PID controller has a transfer function,
\begin{equation}
K_{\rm pid}(s) = k_p+\frac{k_i}{s}+k_ds
\end{equation}
and is therefore completely defined by the three terms $k_p$
(proportional gain),
$k_i$ (integral gain) and $k_d$ (derivative gain). Its frequency
response is
\begin{equation}
K_{\rm pid}(j\omega ) = k_p-j \frac{k_i}{\omega}+jk_d\omega
\end{equation}
Suppose that $K_{\rm pid}(s)$ is used in cascade with an uncertain
plant $G(p,s)$. Then, the nominal open-loop system has frequency response
$L_o(j\omega )=G_o(j\omega)K_{\rm pid}(j\omega)$, where $G_o(s)$ denotes the
nominal plant. Thus, the asymptotic gain of the nominal open loop system
is given by,
\begin{equation}
{\rm lim}_{\omega \rightarrow \infty} \left| G_o (j\omega) \left(
k_p-j \frac{k_i}{\omega}+jk_d\omega \right) \right|
\end{equation}
Suppose that the asymptotic gain of the nominal plant is $|G_o (j\omega ) |
\sim A \omega^{-p}$ where the pole/zero excess of the nominal plant $p$
is at least equal to $2$. Then the asymptotic gain of the nominal open
loop is $| L_o(j\omega ) | \sim A |k_d| \omega^{-p+1}$. Since $A$ and $p$
are fixed, the nominal open-loop gain at high frequencies is minimised
by minimising $|k_d|$. This objective is consistent with the aims of
QFT theory outlined previously.

To design an optimal PID controller, consistent with the requirements
of QFT, we formulate the following optimisation problem:
\blank{1}
Minimise $|k_d|$ subject to the constraints:
\begin{enumerate}
    \item \label{C1}
    $|L_o (j\omega_i)| \geq f_i (\omega_i,\phi)$ for all $i=1,2,...,N$ where
    $\phi := {\rm arg}L_o (j\omega_i)$.
    \item \label{C2}
    $L_o(j\omega_i ) \notin {\cal B}$ for all $i=1,2,...,N$.
\end{enumerate}
Since ${\cal B}$ is a closed contour which is defined for a range of
phases only, it is always possible to combine constraints $(1)$
and $(2)$ for each $i=1,2,...,N$ into a single constraint as long as
$f_i (\omega_i,\phi_i)$ intersects ${\cal B}$ or lies entirely above it, by
calculating the point-wise maximum of the two contours in the common
phase range. This is almost always the case in practice, since robust
performance objectives are almost never associated with frequencies
significantly exceeding the closed-loop bandwidth. The (unlikely)
case that a performance bound lies below the ${\cal B}$ contour can
also be accomodated in our algorithm via an additional checking
condition. To simplify the presentation, however, we will assume in
the sequel that this does not occur, and the combined contours will be
denoted by $\tilde{f}_i(\omega_i, \phi_i)$. The optimisation problem,
therefore, takes the form: Minimise $|k_d|$ subject to
$|L(j\omega_i)| \geq \tilde{f}_i (\omega_i,\phi)$ for all $i=1,2,...,N$.

The magnitude (linear) and phase of $K_{\rm pid}(j\omega )$ are given by,
\begin{equation}
| K_{\rm pid} (j\omega ) | = \sqrt{ k_p^2+ \left( k_d \omega -
\frac{k_i}{\omega} \right) ^2 },\quad {\rm arg} ( K_{\rm pid} (j\omega ) )
:= \psi(\omega ) = {\rm tan}^{-1} \left( \frac {k_d \omega - \frac
{k_i}{\omega}} {k_p} \right)
\end{equation}
respectively. Now suppose that we fix the phase of $K_{\rm pid}$ at
two distinct frequencies $\omega_i$ and $\omega_j$, i.e.
$\psi (j\omega_i)=\psi_i$ and $\psi(j\omega_j)=\psi_j$. Then,
\begin{equation}
\psi(j\omega_i)=\psi_i={\rm tan}^{-1} \left( \frac{k_d \omega_i -\frac{k_i}
{\omega_i}}{k_p} \right)
\end{equation}
which implies that
\begin{equation}
k_d-\frac{k_i}{\omega_i^2}-\frac{k_p {\rm tan}(\psi_i)}{\omega_i}=0
\end{equation}
Similarly,
\begin{equation}
k_d-\frac{k_i}{\omega_j^2}-\frac{k_p {\rm tan}(\psi_j)}{\omega_j}=0
\end{equation}
Equations (8) and (9) can be arranged in matrix form as:
\begin{equation}
\left( \begin{array} {ccc}
1 & - \frac{1}{\omega_i^2} & -\frac{{\rm tan}({\psi_i})}{\omega_i} \\
1 & - \frac{1}{\omega_j^2} & -\frac{{\rm tan}({\psi_j})}{\omega_j}
\end{array} \right)
\left( \begin{array} {c}
k_d \\
k_i \\
k_p
\end{array} \right)
= 0
\end{equation}
Denote the $2 \times 3$ matrix in equation (10) by $A(\psi_i,\psi_j)$.
Clearly, ${\rm Rank}(A(\psi_i,\psi_j))=2$ since $\omega_i \neq
\omega_j$. Therefore, the kernel of $A(\psi_i,\psi_j)$ is a
one-dimensional subspace of ${\cal R}^3$, which implies that the
controller gains $k_d$, $k_i$ and $k_p$ are fixed up to scaling.
Numerically, the kernel of $A(\psi_i,\psi_j)$ can be calculated easily
using the Singular Value Decomposition.

Applying the Singular Value Decomposition to $A(\psi_i,\psi_j)$ gives:
\begin{equation}
A(\psi_i,\psi_j) = \left(
\begin{array} {cc}
U_1 & U_2
\end{array} \right)
\left( \begin{array} {ccc}
\sigma_1 & 0 & 0 \\
0 & \sigma_2 & 0
\end{array} \right)
\left( \begin{array} {c}
V_1^{T} \\
V_2^{T}
\end{array} \right)
\end{equation}
where $V_1^{T}$ is a $2 \times 3$ matrix. Here $V_2$ spans the
kernel of $A(\psi_i,\psi_j)$. Write $V_2^{T}=[ V_{21} ~ V_{22} ~
V_{23} ]$. Then,
\begin{equation}
\left( \begin{array} {c}
k_d \\
k_i \\
k_p
\end{array} \right) = \lambda \left(
\begin{array} {c}
V_{21} \\
V_{22} \\
V_{23}
\end{array} \right)
\end{equation}
where $\lambda$ is an arbitrary real constant. Using equation (6),
the gain and phase of $K_{\rm pid}(s)$ may be written as
\begin{equation}
| K_{\rm pid} (j\omega ) | = | \lambda | \sqrt{ V_{23}^2+ \left( V_{21}
\omega - \frac{V_{22}}{\omega} \right) ^2 }, \quad
\psi(\omega ) = {\rm tan}^{-1} \left( \frac {V_{21} \omega - \frac
{V_{22} }{\omega}} {V_{23}} \right)
\end{equation}
Note that equation (13) implies that fixing the phase of
$K_{\rm pid}(j\omega )$ at two frequencies, fixes the phase
of $K_{\rm pid}(j\omega )$ at {\it any} frequency $\omega$, and thus also
the phase of $L_o(j\omega )$. In this case, minimising $|k_d |$ is
equivalent to minimising $| \lambda V_{21} |$.

Under the constraint that $\psi(\omega_i)=\psi_i$ and $\psi(\omega_j)=
\psi_j$, the QFT constraints are satisfied if and only if
\begin{equation}
| L_o(j\omega_k) |_{\rm db} \geq \tilde{f}_k (\omega_k,\phi_k)
\end{equation}
for all $k=1,2,...,N$, where $\phi_k={\rm arg}(L_o(j\omega_k))$.
Note that since the phase of $K_{\rm pid}(j\omega)$ is fixed at every
frequency $\omega$, $\phi_k$ is fixed and {\it known}
for all $k=1,2,...,N$. In fact,
\begin{equation}
\phi_k = {\rm arg}(G_o(j\omega_k))+{\rm tan}^{-1} \left( \frac
{V_{21}\omega_k - \frac{V_{22}}{\omega_k}}{V_{23}} \right)
\end{equation}
Since,
\begin{equation}
|L_o(j\omega_k)|_{\rm db} = |G_o(j\omega_k)|_{db}+|K_{\rm pid}(j\omega_k)|_
{\rm db}
\end{equation}
equation (14) is equivalent to
\begin{equation}
|K_{\rm pid}(j\omega_k)|_{\rm db} \geq \tilde{f}_k(\omega_k, \phi_k)-
|G_o(j\omega_k )|_{\rm db}
\end{equation}
for all $k=1,2,...N$. Substituting from (13) shows that this is
equivalent to
\begin{equation}
20 {\rm log}_{10} | \lambda | \geq {\rm max}_{k=1,2,...,N}
\left( \tilde{f}_k (\omega_k, \psi_k ) - |G_o(j\omega_k ) |_{\rm db}
-10 {\rm log}_{10} \left( V_{23}^2 + \left(V_{21}\omega_k -\frac {V_{22}}
{\omega_k} \right) ^2 \right) \right)
\end{equation}
or
$ | \lambda | \geq 10^{\frac{\beta}{20}}$
where we have defined,
\begin{equation}
\beta = {\rm max}_{k \in \{ 1,2,...N\} } \left( \tilde{f}_k (\omega_k,
\phi_k) - |G_o(j\omega_k) |_{\rm db} -10 {\rm log_{10}} \left(
V_{23}^{2}+\left(V_{21}\omega_k - \frac{V_{22}}{\omega_k} \right) ^2
\right) \right)
\end{equation}
Multiplying by $| V_{21} |$ and noting that $| k_d |=| \lambda
V_{21} |$, implies that $|k_d| \geq | V_{21} | 10^{ \frac{\beta}{20}}$.
Hence, provided that the phase of $K_{\rm pid}$ is fixed
as ${\rm arg} ( K_{\rm pid}(j\omega_i) ) =\psi_i$ and
${\rm arg} ( K_{\rm pid}(j\omega_j) ) =\psi_j$, the minimum
value of $|k_d |$ which achieves the robust-performance constraints is
given by $|k_d^{*}|=|V_{21}^{i,j}10^{\frac{\beta_{i,j}}{20}}|$, where
the additional indexes $(i,j)$ introduced in $V_{21}$ and $\beta$ emphasise the
dependence of these variables on $(\omega_i,\omega_j)$ and $(\psi_i,\psi_j)$.
We can now formulate the following algorithm for solving the optimisation
problem. In this algorithm, the gains
$k_i$, $k_d$ and $k_p$ have been further constrained to be non-negative.
This assumption can be removed, if desired,  with minor modifications to the
algorithm.

{\bf Algorithm 1:} Given the plant's uncertainty templates ${\cal G}_i$,
a nominal plant $G_o(j\omega_i) \in {\cal G}_i$ and QFT contraint
bounds $\tilde{f}_i ( \omega_i,\phi)$, each defined at frequencies
$\omega_1,\omega_2, ... , \omega_N$, the following algorithm calculates
an optimal PID controller $K_{\rm pid} ^{*}(s) =
k_p^{*}+\frac{k_i^{*}}{s}+k_d^{*}s$ with non-negative gains,
if it exists:

\begin{enumerate}
\item Obtain a phase array $\Phi$ by discretising the phase interval
$(-360^{\circ},0^{\circ})$.
\item Select any two distinct frequencies $\omega_k,\omega_l \in \{ \omega_1,
\omega_2 , ..., \omega_N \}$.
\item Caclulate phase intervals $\Phi_k,\Phi_l \subseteq \Phi$ in which the nominal
open-loop phase can vary at $\omega_k$ and $\omega_l$ if a PID controller
is used. $\Phi_k$ ($\Phi_l$) contains
the phases of $\Phi$ which lie within $\pm 90^{\circ}$ of the nominal plant phase
$G_o(j\omega_k)$ ($G_o(\omega_l)$).
\item Initialise $m \times n$ arrays $K_p$, $K_i$ and $K_d$, where
$m$ and $n$ are the sizes of $\Phi_k$ and $\Phi_l$ respectively.
\item For each $(\Phi_k(i),\Phi_l(j)) \in \Phi_k \times \Phi_l$
\begin{enumerate}
\item Calculate $\psi_i = \Phi_k(i)-{\rm arg}(G_o(\omega_k))$ and
$\psi_j = \Phi_l(j)-{\rm arg}(G_o(\omega_l))$.
\item Calculate the singular value decomposition of $A(\psi_i,\psi_k)$ and the
corresponding vector $V_2^{i,j}$ spanning its kernel.
\item If any two elements of $V_2^{i,j}$ have opposite signs, set $K_d (i,j)=\infty$;
Else, let $q$ be the (common) sign of the elements of $V_2^{i,j}$, calculate
$\beta_{i,j}$ using equation (17), and set $\lambda = q 10^{\frac {\beta_{i,j}}
{20}}$, $K_d(i,j)=\lambda V_{21}^{i,j}$, $K_i(i,j)=\lambda V_{22}^{i,j}$ and
$K_p(i,j)=\lambda V_{23}^{i,j}$.
\end{enumerate}
\item Calculate $(i^{*},j^{*}) \in {\rm argmin}_{i,j} K_d(i,j)$. If
$K_d(i^{*},j^{*}) = \infty$, the constraints cannot be satisfied by a PID
controller with non-negative gains; otherwise the optimal PID controller is
$K^{*}(s)=K_p(i^{*},j^{*})+K_i(i^{*},j^{*})s^{-1}+K_d(i^{*},j^{*})s$.
\end{enumerate}

{\bf Remarks on Algorithm 1:}

\begin{itemize}
\item In step 1 the phase discretisation of the interval
$(-360^{\circ},0^{\circ})$ results
in a grid of phases, typically equally-spaced. In practice, $50$-$100$
phases are adequate. It is helpful to calculate the
performance bounds over the same phase grid ($\Phi$).
\item In principle any two frequencies $\omega_k$, $\omega_l$ can be selected
in step 2. Selecting these two frequencies reasonably far apart works well
in practice.
\item In step 3, the phase intervals $\Phi_k$ and $\Phi_l$ may be further restricted,
if desired, to ensure that the nominal open-loop frequency response $L_o(j\omega)$
is shaped appropriately. This will also reduce the number of calculations
in step 5 of the algorithm.
\item The singular value decomposition in step 5(b) of the algorithm can be
dispensed with altogether, by calculating $V_2^{i,j}$ analytically. This,
however, increases the complexity of the algorithm and does not lead to any
significant reduction in computation time.
\item Step 5(c) of the algorithm requires the calculation of $\beta_{i,j}$
which in turn relies on the calculation of the performance bounds at phases
specified by equation (15), which may not belong to the (discrete) phase array used
to calculate the bounds (typically $\Phi$).
There is no difficulty, however, in estimating
the performance bounds for these phases using the adjacent phase points,
e.g. via linear interpolation. Alternatively, the performance bounds can be
calculated exactly for the phases obtained from equation (15) via a bisection
algorithm implemented between steps 5(a) and 5(c).
\item The algorithm can be specialised, if desired, to calculate optimal
PI of PD controllers. This requires only a one-dimensional search over a
phase grid defined at a single frequency.
\end{itemize}

\section{Example}

In this section we illustrate our algorithm by means of a simple example.
The uncertain plant is taken as
\begin{equation}
G(s)=\frac{k a}{s^2+as}
\end{equation}
in which the parameters $a$ and $k$ vary independently in the intervals
$1 \leq a \leq 10$ and $1 \leq k \leq 10$ respectively. The nominal plant,
$G_o(s)$, is taken to correspond to $a=1$ and $k=1$. The tracking
specifications are defined as:
\begin{equation}
|B_l(j\omega_i)| \leq |T(a,k,j\omega_i)| \leq |B_u(j\omega_i)|
\end{equation}
Here $T(a,k,s)=F(s)G(a,k,s)K(s)(1+G(a,k,s)K(s))^{-1}$ is the closed-loop
transfer function and the lower and upper tracking bounds are defined
as the magnitude frequency responses of the two systems
\begin{equation}
B_l(s)=\frac{0.6585(s+30)}{(s+2+j3.969)(s+2-j3.969)}
\end{equation}
and
\begin{equation}
B_u(s)=\frac{8400}{(s+3)(s+4)(s+10)(s+70)}
\end{equation}
at each $s=j\omega_i$.
Note that the zero of $B_l(s)$ ($s=-30$) and the two fast poles in
$B_u(s)$ ($s=-10$ and $s=-70$) have been included to ensure that the
the magnitude frequency responses of $B_l(s)$ and $B_u(s)$ diverge
at high frequencies \cite{DAH}. The frequencies of interest
$\{\omega_i \}$ have been selected as $\omega_1=0.5$, $\omega_2=1$,
$\omega_3=2$, $\omega_4=3$, $\omega_5=5$, $\omega_6=10$, $\omega_7=30$ and
$\omega_8=60$ rads/s. For simplicity, no disturbance-rejection objectives
have been considered in this example.

The uncertainty templates of the plant at the eight frequencies of interest
have been calculated numerically and are displayed in Figure 2. To reduce the
number of calculations, each template has been replaced by its convex hull.
This results in a minimal amount of conservativeness in this case.

\begin{figure}[htb]
   \epsfig{file=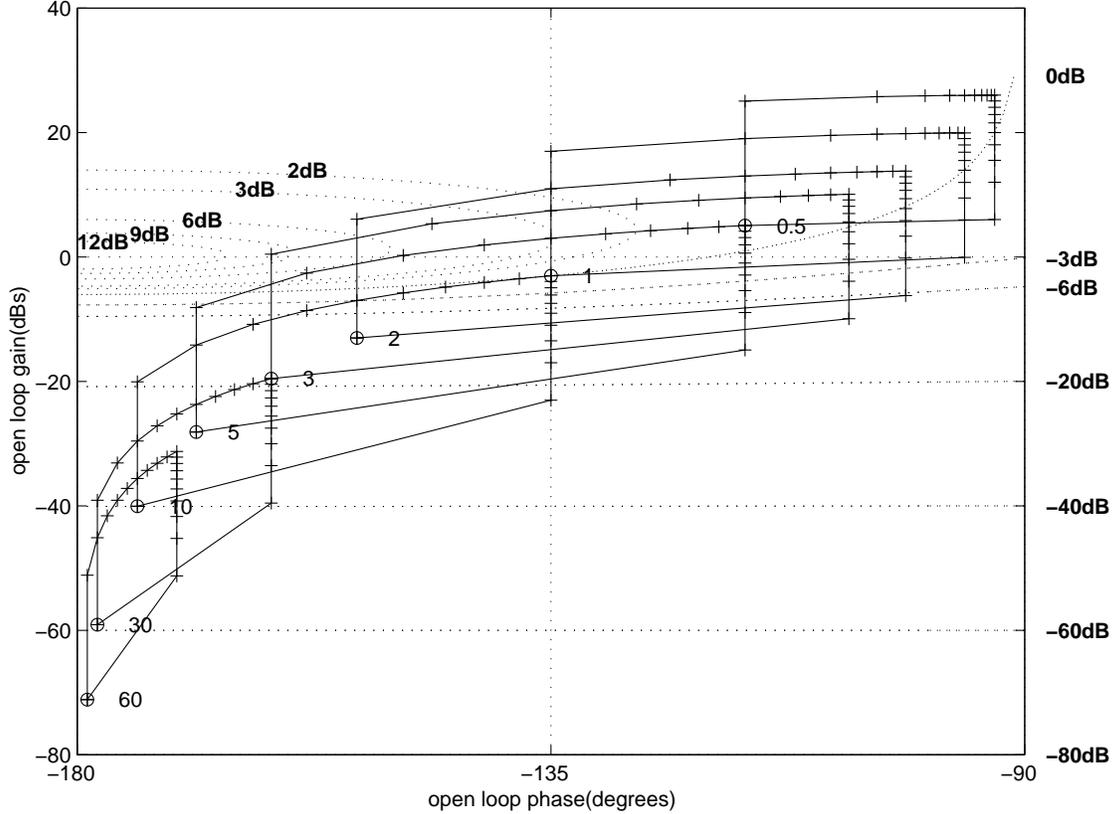,scale=0.8}
   \caption{Uncertainty templates}
   \label{fig:2}
\end{figure}

Next, an optimal PID controller was designed following the procedure of
Algorithm 1. The optimal controller was obtained as
\begin{equation}
K_{\rm pid}^{*}(s)=12.6+3.95s+\frac{4.46}{s}
\end{equation}

\begin{figure}[htb]
   \epsfig{file=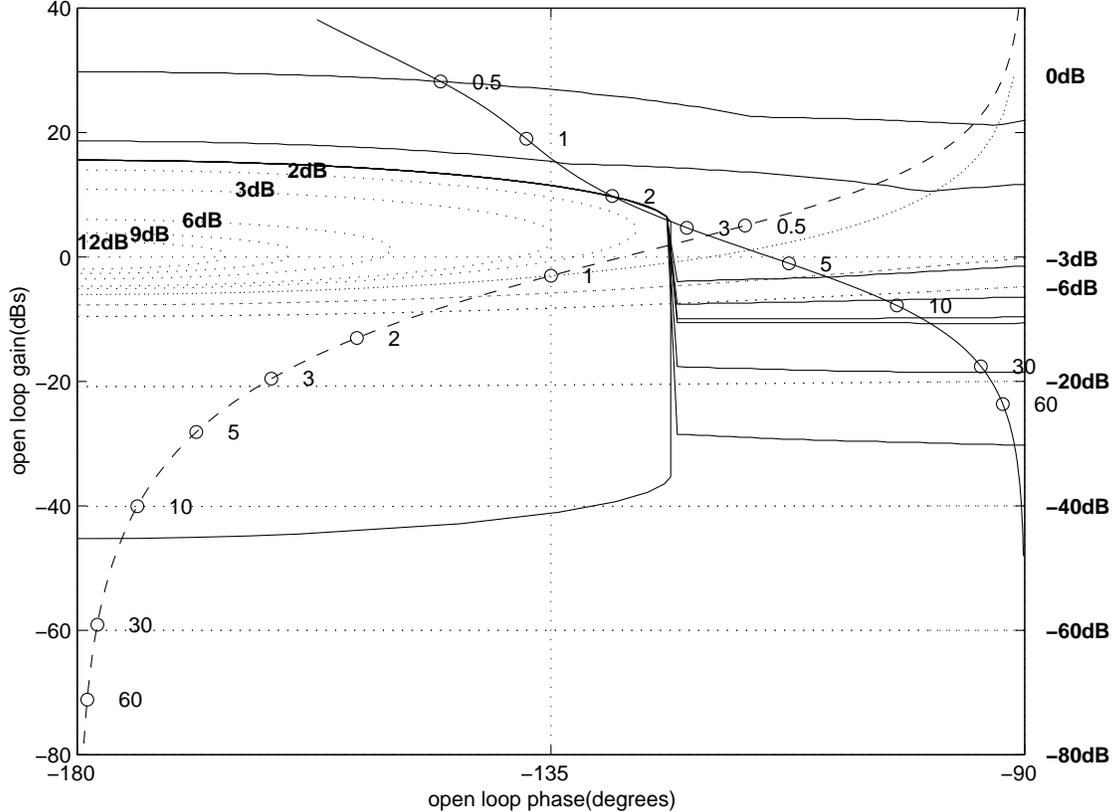,scale=0.8}
   \caption{Nominal plant and nominal open-loop}
   \label{fig:3}
\end{figure}

Figure 3 shows the frequency response of the nominal plant
(dashed line) and the nominal open loop (solid line) on the
Nichols chart, together with the eight Horowitz templates and the
${\cal B}$-contour (corresponding to an M value of $1.2$). The
eight frequencies of interest are indicated by circles on the two
frequency responses. The design meets the specifications, since
the nominal frequency response does not penetrate the ${\cal
B}$-contour and lies on or above the Horowitz templates at the
eight frequencies of interest. As expected, one of the
$L(j\omega_i)$'s ($L(j\omega_4)$) lies exactly on a bound (in
this case the ${\cal B}$ contour).

Since the open-loop system has a pole-zero excess equal
to one, its phase approaches $-90^{\circ}$ at high frequencies. There is no
difficulty, however, in forcing the open loop response to approach the $-180^{\circ}$
phase line at high frequencies, if desired, by including a suitable
$1+s \tau$ factor in the denominator of the controller derivative term. Choosing,
for example, $\tau \ll \omega_N^{-1}$, has a minimal effect on the shape
of $L_o(j\omega)$ for $\omega \leq \omega_N$. Alternatively, the PID controller
may be assumed to be of the form
\begin{equation}
K_{\rm pid}^{\prime}(s)=k_p+\frac{k_i}{s}+\frac{k_d s}{1+s \tau}
\end{equation}
($\tau$ fixed) before solving the optimisation problem. Since in this case
\begin{equation}
L_o^{\prime}(p,s):=K_{\rm pid}^{\prime}(s)G(p,s)=\left( (k_p+k_i\tau)+
\frac{k_i}{s}+( k_d+k_p\tau )s \right) \frac{G(p,s)}{1+s\tau}
\end{equation}
our algorithm can still be applied by redefining the uncertain plant as
\begin{equation}
G^{\prime}(p,s)=\frac{G(p,s)}{1+s\tau}
\end{equation}
and optimising with respect to the new variables $k_p^{\prime}=k_p+k_i\tau$,
$k_i^{\prime}=k_i$ and $k_d^{\prime}=k_d+k_p\tau$.

A second-order prefilter $F(s)$ (of dc gain equal to $1$ and cut-off frequencies
$3.5$ and $7.5$ rads/s) was finally designed using the standard procedure \cite{DAH}.
Figure 4 shows the closed-loop frequency responses for a number of $(a,k)$ parameter
combinations, together with the tracking bounds $|B_l(j\omega)|$ and $|B_u(j\omega)|$.
Again, the eight frequencies of interest are marked by circles.
Clearly, the specifications of the design are met, as expected from the
characteristics of the open loop response in Figure 3.

\begin{figure}[htb]
   \epsfig{file=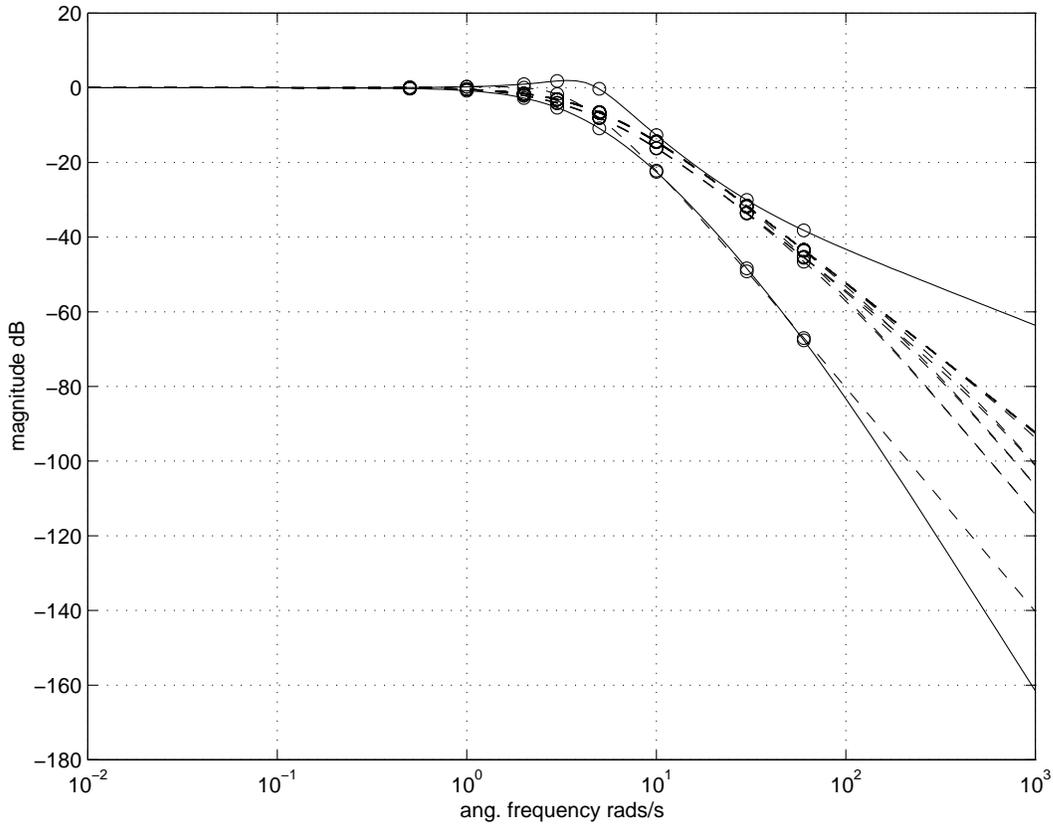,scale=0.8}
   \caption{Closed-loop responses and bounds}
   \label{fig:4}
\end{figure}

\section{Conclusions}

An algorithm has been presented for designing optimal PID controllers for
uncertain systems subject to QFT constraints. The algorithm is simple, easy
to implement and can be used to automate the loop-shaping step of the QFT
design procedure. Although the algorithm has been presented for SISO systems,
its extension to multivariable problems is possible using the standard QFT
approach.

\section{Acknowledgement}
A. Zolotas would like to thank Mr Dimitriadis and Sevath ABE for providing
him with the necessary computer equipment during the last phase of this work.

\end{document}